\documentclass[aps,prl,twocolumn,nofootinbib,longbibliography,,superscriptaddress,10pt]{revtex4-2}
\usepackage{etoolbox}
\usepackage{dcolumn,lipsum}
\usepackage{amsmath,amssymb,amsfonts,mathtools}
\usepackage{mathrsfs,bbold}
\usepackage{graphicx}
\usepackage[colorlinks=true,urlcolor=blue,citecolor=red,linkcolor=blue]{hyperref}
\usepackage{accents}
\newlength{\dhatheight}
\usepackage{natbib}
\usepackage{wrapfig}
\usepackage{flushend,BOONDOX-cal,BOONDOX-frak}

\makeatletter
\newsavebox{\@brx}
\newcommand{\llangle}[1][]{\savebox{\@brx}{\(\m@th{#1\langle}\)}%
  \mathopen{\copy\@brx\kern-0.5\wd\@brx\usebox{\@brx}}}
\newcommand{\rrangle}[1][]{\savebox{\@brx}{\(\m@th{#1\rangle}\)}%
  \mathclose{\copy\@brx\kern-0.5\wd\@brx\usebox{\@brx}}}
\makeatother

\begin{document}
\title{Towards graviton lasing from squeezed ultra-cold systems}
\author{Soham Sen}
\email{sensohomhary@gmail.com}
\affiliation{Department of Astrophysics and High Energy Physics, S. N. Bose National Centre for Basic Sciences, JD Block, Sector-III, Salt Lake City, Kolkata-700 106, India}
\author{Vlatko Vedral}
\email{vlatko.vedral@physics.ox.ac.uk}
\affiliation{Clarendon Laboratory, University of Oxford, Park Road, Oxford OX1 3PU, United Kingdom}

\begin{abstract}
\noindent 
In our recent work, \href{https://arxiv.org/abs/2604.11474}{arXiv:2604.11474 [hep-th]}, we have shown that effective detection of gravitons is possible using an array of charged harmonic oscillators in a dynamical electromagnetic field. Using the interaction Hamiltonian of the identical model, we find out that a systematic way of population inversion of the gravitons is possible in ultra-cold atomic systems. We find out that the exponential growth depends strictly on the number of bosons in the system as well as their inherent squeezing of the matter wave packets. A coherent source of gravitons may lead directly to an unavoidable evidence on the existence of gravitons and based on this analysis we propose an experimental proposal for generating true graviton laser.
\end{abstract}
\maketitle
\textit{Introduction-} In low energy quantum gravity theory, graviton is considered to be the quanta of gravitational fluctuations working as a mediator of the force. The primary aim in such works is to look for the response of quantum matter towards incoming quantum gravitational fluctuations \cite{MarlettoVedral2,
EntanglementQuantumGravity,
EntanglementQuantumGravity2,
EntanglementQuantumGravity3,
QGravLett,QGravD,KannoSodaTokuda,KannoSodaTokuda2}.  The two mainstream series of works focus primarily on two aspects of gravitons, one being the graviton induced noise on the quantum matter and the other being the entangling power of gravity where the gravitons act as the mediator of entanglement. Another way of detecting gravitons would be to look for spontaneous emission of gravitons which is however argued to be not possible to detect using any realistic experimental set-up by Freeman Dyson \cite{Dyson}. In a recent work \cite{ORM}, we have investigated a very simple model of charged harmonic oscillator, where the system is placed inside of an electromagnetically shielded cavity where the cavity is filled with electromagnetic radiation. Starting from the action of the model and deriving the Hamiltonian for the system, we find out that the system allows a trilinear interaction term which couples the graviton degrees of freedom with the detector as well as the photon. This interaction is fundamentally and physically different from the Gertsenshtein term \cite{Gertsenshtein}. The trilinear interaction allows for three independent physical processes to occur depending on the resonance condition where we have primarily focussed on the case where the graviton frequency is equal to the sum of the frequency of the detector and the photon. In such a scenario, if a high frequency graviton is absorbed by the detector then it jumps to higher excited state while simultaneously emitting a photon. The reverse process actually allows for spontaneous as well as stimulated emission of gravitons. We find out that if the cavity is pumped with photons with identical frequency the stimulated as well as the spontaneous emission probability can be increased significantly. We, however, have actually ignored the other two physical processes one of which will be the primary matter of focus in our current manuscript.
 
\noindent If there are more quanta in the excited quantum states than the lower energy level, then the spontaneously emitted photons can initiate a stimulation process where a beam of coherent photons are emitted. This coherent beam of stimulated photons are known as light amplification by stimulated emission of radiation or a LASER. In case of gravitons there is a proposal of creating graviton laser by using ultra-cold neutrons which also had some caveats related to physical realism and experimental implementation \cite{Graviton_Laser}. In our work, using the model used in \cite{ORM} and exploiting the terms of the trilinear interaction, we propose a realistic analytical scenario where it will be possible to create a true graviton laser. For our model, we show that using Bose-Einstein condensates with squeezed phonon modes, we can obtain a genuine physical population inversion scenario and based on this theoretical model, we propose a real graviton laser model. The graviton laser will be the best possible way for detecting the existence of gravitons as it allows for the detection of a coherent and collective graviton beam resulting in an amplified spacetime distortion than will be observable for a single graviton detection scenarios. In this work, we start with the Hamiltonian from our previous model and derive the master equation. From there, we obtain the time evolution equation for the graviton number and investigate the conditions for population inversion. Finally, we propose an experimental model and conclude our results.

\textit{The model-}
\noindent In \cite{ORM}, we have considered an array of identical charged harmonic oscillators in a cavity QED set-up in the presence of a gravitational wave background where the electromagnetic field is considered to be dynamical in nature. The action for the model system in the gravitational wave background ($g_{\mu\nu}=\eta_{\mu\nu}+h_{\mu\nu}$) was considered to be $S
=\frac{1}{16\pi G}\int d^4 x\sqrt{-g}R-\frac{1}{4}\int d^4x \sqrt{-g}F_{\mu\nu}F^{\mu\nu}-m_0\int dt \left[\sqrt{-g_{\mu\nu}\dot{\mathcal{Y}}^\mu\dot{\mathcal{Y}}^\nu}+\frac{\omega_0^2}{2} \mathcal{Y}_\mu\mathcal{Y}^\mu-\frac{q}{m_0} g_{\mu\nu}A^\mu \dot{\mathcal{Y}}^\nu\right]$ with $\mathcal{Y}^\mu=\{t,\xi^i\}$ denoting the Fermi-normal coordinated for the oscillators, $\omega_0$ being the oscillator frequency, $q$ being the charge carried by each oscillator, and $m_0$ denoting the mass of the detector. Now, we use the transverse-traceless gauge condition
$h^{\text{TT}}_{\mu\nu}(t,\vec{x})\rightarrow\bar{h}_{ij}(t,\vec{x})=\frac{1}{\sqrt{\hbar G}}\sum_{\vec{k},s}h_{s}(t,\vec{k})e^{i\vec{k}\cdot\vec{x}}\epsilon_{ij}^s(\vec{k})$
where $h^s(t,\vec{k})$ is the Fourier mode functions, $\epsilon^s_{ij}(\vec{k})$ is the polarization tensor, and $s=\{+,\times\}$ denotes the polarization for the gravitational fluctuations. We also make use of the coulomb gauge condition ($A_0=0,$ $\vec{\nabla}\cdot\vec{A}=0$) and write the Fourier mode decomposition of the vector field as $A_{i}(t,\vec{x})=\frac{1}{\sqrt{\hbar G^2}}\sum_{\vec{k}_P,P}A_{P}(t,\vec{k}_P)e^{i\vec{k}_P\cdot\vec{x}}\epsilon^s_{i}(\vec{k}_P)$
where $A_{P}(t,\vec{k}_P)$ denotes the Fourier mode function in the Fourier space and $\epsilon^s_i(\vec{k}_P)$ denotes the electromagnetic polarization tensor. 
As has been discussed in \cite{ORM}, we can set the reality condition for the gravitational as well as the electromagnetic field as $\bar{h}_{ij}(t,\vec{x})=\bar{h}^*_{ij}(t,\vec{x})$ and $A_{i}(t,\vec{x})=A^*_i(t,\vec{x})$. For single mode consideration and plus polarization for both the fields, we can define $h(t)\equiv\Re[h_+(t,k)]$ and $A(t)\equiv\Re[A_{+}(t,k_P)]$. This helps us to read off the Lagrangian from the action and write it in a more simple form. The Hamiltonian then is obtained from the Lagrangian after quantization as \cite{ORM}
\begin{equation}\label{II.2A}
\begin{split}
\hat{H}\simeq&\hat{H}_0+\frac{2g_P h}{m_P}\left(\frac{\hat{p}_A^2}{2m_P}-\frac{1}{2}m_p\omega_P^2\hat{A}^2\right)\\
+&\frac{g_h}{2mm_0}\hat{p}_h(\hat{\xi}\hat{\pi}_\xi+\hat{\pi}_\xi \hat{\xi})-\frac{q_P}{m_0}\hat{A}\hat{\pi}_\xi-\frac{g_hq_P}{mm_0}\hat{p}_h\hat{A}\hat{\xi}~.
\end{split}
\end{equation}

\textit{The master equation-}
The trilinear interaction Hamiltonian for a charged harmonic oscillator in the presence of a dynamical electromagnetic field and background gravitational perturbation reads ($c$ is set to unity) 
\begin{equation}\label{II.1}
\hat{H}_{\text{int}}=-\frac{g_hq_p}{mm_0}\hat{p}_h\otimes \hat{A}\otimes\hat{\xi}
\end{equation}
where $g_h\equiv\frac{m_0}{2\sqrt{\hbar G}}$, $q_p\equiv=\frac{q}{\sqrt{\hbar G^2}}$, and $m\equiv \frac{V}{16\pi \hbar G^2}$ with $V$ denoting the quantization volume, $m_0$ denoting the mass of the harmonic oscillator, and $q$ denoting its charge. Here, $\{\hat{h},\hat{p}_h\}$ denotes the phase space operators of the graviton mode, $\{\hat{A},\hat{p}_A\}$ denotes the photon phase-space operators, and $\{\hat{\xi},\hat{\pi}_\xi\}$ denotes the phase space operators corresponding to the detector.
Here we are working in the mostly positive signature. In order to derive the master equation, one needs to write down the interaction Hamiltonian in the interaction picture as
$\hat{\mathcal{H}}(t)\equiv\hat{H}^I_{int}(t)=e^{\frac{i}{\hbar}\hat{H}_0 t}\hat{H}_{int}e^{-\frac{i}{\hbar}\hat{H}_0 t}$ with $\hat{H}_0$ being the base Hamiltonian for the model system. In the interaction picture, the interaction Hamiltonian in eq.(\ref{II.1}) takes the form $\hat{\mathcal{H}}(t)=-\frac{g_hq_P}{mm_0}\hat{p}_h^I\otimes \hat{A}^I\otimes\hat{\xi}^I$ where the phase space operators in the interaction picture are given as $\hat{p}_h^I=i\sqrt{\frac{m\omega\hbar}{2}}(\hat{b}^\dagger e^{i\omega t}-\hat{b}e^{-i\omega t})$, $\hat{A}^I=\sqrt{\frac{\hbar}{2m_P\omega_P}}(\hat{a}e^{-i\omega_P t}+\hat{a}^\dagger e^{i\omega_Pt})$, and $\hat{\xi}^I=\sqrt{\frac{\hbar}{2m_0\omega_0}}(\hat{\chi}e^{-i\omega_0 t}+\hat{\chi}^\dagger e^{i\omega_0 t})$. Here, $\omega$, $\omega_P$, and $\omega_0$ denote the graviton, photon and detector frequencies respectively. In the above expressions, the ladder operators satisfy $[\hat{b},\hat{b}^\dagger]=[\hat{a},\hat{a}^\dagger]=[\hat{\chi},\hat{\chi}^\dagger]=1$.
We now start by considering the total density matrix of the system to be $\hat{\rho}(t)$ which obeys $i\hbar\dot{\hat{\rho}}(t)=[\hat{H},\hat{\rho}(t)]$ where $\hat{H}=\hat{H}_0+\hat{H}_{\text{int}}$. In the interaction picture, the density matrix is expressed as $\hat{\rho}^I(t)=e^{\frac{i}{\hbar}\hat{H}_0t}\hat{\rho}(t)e^{-\frac{i}{\hbar}\hat{H}_0t}$ and the von-Neumann equation takes the form $\frac{d\hat{\rho}^I(t)}{dt}=-\frac{i}{\hbar}\left[\hat{\mathcal{H}}(t),\hat{\rho}^I(t)\right]$. Replacing the solution of $\hat{\rho}^I(t)$ in the von-Neumann equation, we arrive at the well-known form as 
$\frac{d\hat{\rho}^I(t)}{dt}=-\frac{i}{\hbar}[\hat{\mathcal{H}}(t),\hat{\rho}^I(0)]-\frac{1}{\hbar^2}\int_0^t dt'[\hat{\mathcal{H}}(t),[\hat{\mathcal{H}}(t'),\hat{\rho}^I(t')]]$.
As we are interested in the time evolution of the graviton number operators, we proceed with the calculation of the time-evolution of the reduced density matrix which is obtained by taking the trace over the photon and detector degrees of freedom as $\dot{\hat{\rho}}^I_{\text{GR}}(t)=\text{tr}_{\text{D}}[\text{tr}_\text{P}[\dot{\hat{\rho}}^I(t)]$. The trace-reduced equation of motion takes the form \footnote{The first order commutator vanishes after taking the trace as the photon operators appear in the linear order in the interaction Hamiltonian.}
\begin{equation}\label{II.2}
\frac{d\hat{\rho}^I_{\text{GR}}(t)}{dt}=-\frac{1}{\hbar^2}\int_0^tdt' \text{tr}_{\text{D}}\left[\text{tr}_{\text{P}}\left[[\hat{\mathcal{H}}(t),[\hat{\mathcal{H}}(t'),\hat{\rho}^I(t)]]\right]\right]
\end{equation}
where we have implemented the Markov approximation by replacing $\hat{\rho}^I(t')$ by $\hat{\rho}^I(t)$. The interaction Hamiltonian in the interaction picture can now be divided into three independent sectors based on the three possible resonant conditions. Here, we are interested in the physical process where $\omega_0=\omega+\omega_P$ gets satisfied. In such a case when the detector absorbs one graviton with energy $\hbar \omega$ and one photon with energy $\hbar\omega_P$, it jumps one energy level. If it now suddenly jumps down to its lower excited state, it simultaneously emits one photon and one graviton. 
 If we now define $\mathcal{A}\equiv \frac{g_h q_p}{2mm_0}\sqrt{\frac{m\hbar\omega}{2m_P\omega_Pm_0\omega_0}}$, we can write down this sector of the interaction Hamiltonian in the interaction picture as
 \begin{equation}\label{II.3} 
 \hat{\mathcal{H}}^{GL}(t)=-i\hbar \mathcal{A}\left
 (\hat{b}^\dagger\hat{a}^\dagger\hat{\chi}\exp[i\varpi t]-\hat{b}\hat{a}\hat{\chi}^\dagger \exp[-i\varpi t]\right)
 \end{equation}
where the effective frequency is defined as $\varpi\equiv \omega+\omega_P-\omega_0$. We now need to implement the Born approximation by writing the density matrix in the interaction picture as $\hat{\rho}^{I}(t)$ and $\hat{\rho}^I(t)=\hat{\rho}^I_{\text{GR}}(t)\otimes \hat{\rho}_P\otimes \hat{\rho}_{\text{D}}$. Here, both photons as well as the detector degrees of freedom are considered as bath. One can also proceed by considering $\hat{\rho}_P$ to be time dependent. 

\textit{Generating population inversion:} We consider a standard system for which the detector part of the density matrix reads $\hat{\rho}_{\text{D}}=|\mathcal{N}\rangle\langle \mathcal{N}|$ and the photon density matrix reads $\hat{\rho}_P=|n_P\rangle\langle n_P|$. Here, $\hat{a}|n_P\rangle=\sqrt{n_P}|n_P-1\rangle$. Depending on the model $\mathcal{N}$ may denote the energy level for the oscillating system and in special cases can also indicate the number of phonons in the quantum state. Here $\mathcal{N}$ is simply given by $\mathcal{N}=\text{tr}_\text{D}[\hat{\rho}_{\text{D}}\hat{\chi}^\dagger\hat{\chi}]$. It is then easy to understand that  In such a scenario the Lindblad master equation takes the form 
\begin{equation}\label{II.4}
\begin{split}
2\dot{\hat{\rho}}^I_{\text{GR}}=&-\Gamma_{\text{GL}}\left[(n_P+1)\mathcal{N}\left[2\hat{b}^\dagger \hat{\rho}^I_{\text{GR}}\hat{b}-\{\hat{b}\hat{b}^\dagger,\hat{\rho}^I_{\text{GR}}\}\right]\right]\\
&-\Gamma_{\text{GL}}\left[n_P(\mathcal{N}+1)\left[2\hat{b} \hat{\rho}^I_{\text{GR}}\hat{b}^\dagger-\{\hat{b}^\dagger\hat{b},\hat{\rho}^I_{\text{GR}}\}\right]\right]
\end{split}
\end{equation}
where $\Gamma_{\text{GL}}\equiv2\pi \mathcal{A}^2\delta(\omega +\omega_P-\omega_0)$. If we now use the number operator for the gravitons then the time evolution equation for $n_G(t)=\langle \hat{b}^\dagger\hat{b}~\hat{\rho}^I_{\text{GR}}(t)\rangle$, as\footnote{For a detailed derivation of the master equation please see the end matter.}
\begin{equation}\label{II.5}
\frac{dn_G(t)}{dt}=\Gamma_{\text{GL}} (\mathcal{N}-n_P)~n_G(t)+\Gamma_{\text{GL}}\mathcal{N}(n_P+1)~.
\end{equation}
It is evident from the above equation that for population inversion to occur $\mathcal{N}>n_P$ must hold true. For a simple resonant detector model as has also been discussed in \cite{ORM}, the detector state denotes the energy level for the model and hence it is quite difficult to achieve population inversion just based on the depletion of the number of photons inside of the cavity. Even for systems where the phonon operators couple to the graviton as well as the photons, high phonon pumping cannot lead to just two level population inversion as has been discussed in \cite{Graviton_Laser} (using pumping of ultra-cold neutrons). 
It is quite evident that even for systems with high phonon numbers, it is quite difficult to obtain the above population inversion condition. We therefore proceed towards considering ultra cold atomic systems as the base of our analysis. 

\textit{Ultra-cold atoms for generating population inversion:} In case of a Bose-Einstein condensate the state will be replaced by coherent state and we consider squeezed phonons which allows for a highly controllable experimental parameter. The squeezing parameter is now given by $\zeta=r \exp[i\phi]$ where $r$ is the squeezing parameter and $\phi$ is the squeezing angle. The squeezed coherent state is then represented by $|\mathcal{N}\rangle=\hat{S}_\zeta|\chi\rangle$ such that $\hat{\chi}|\chi\rangle=\chi|\chi\rangle$ where $\chi=\sqrt{N}e^{i\theta}$ with $N$ being the number of atoms in the condensate and $\theta$ denoting the associated phase. The raising and lowering operators under the action of the squeezing operator transforms as
$\hat{S}_\zeta^\dagger \hat{\chi}\hat{S}_\zeta=\hat{\chi} \cosh r-\hat{\chi}^\dagger e^{i\phi}\sinh r$ and $\hat{S}_\zeta^\dagger \hat{\chi}^\dagger\hat{S}_\zeta=\hat{\chi}^\dagger \cosh r-\hat{\chi} e^{-i\phi}\sinh r$. One can then obtain the analytical expression of $\mathcal{N}=\text{tr}_{\text{BEC}}\left[\hat{\rho}_{\text{BEC}}\hat{\chi}^\dagger\hat{\chi}\right]$ as $\text{tr}_{\text{BEC}}\left[\hat{\rho}_{\text{BEC}}\hat{\chi}^\dagger\hat{\chi}\right]=N\cosh 2r-N\cos\varphi\sinh 2r+\sinh^2 r$ where $\varphi\equiv 2\theta-\phi$ and $|\chi|^2=N$. For high phonon squeezing, the time evolution equation for the population of gravitons from eq.(\ref{II.5}) takes the form
\begin{equation}\label{II.6}
\begin{split}
\frac{dn_G(t)}{dt}=&\Gamma_{\text{GL}}\left[e^{2r}N_\varphi-n_P\right]n_G(t)+\Gamma_{\text{GL}}e^{2r}N_\varphi(n_P+1)
\end{split}
\end{equation}
where the relative angle $\varphi$ dependent $N_\varphi$ is defined as $N_\varphi\equiv \frac{N}{2}(1-\cos\varphi)+\frac{1}{4}$.
It is now important to analyze eq.(\ref{II.6}) thoroughly. It is evident that for population inversion to occur, $e^{2r}N_\varphi > n_P$ condition must hold true. Few important observations are now in order. If the squeezing angle is tuned in a way such that $\varphi=2n \pi$ with $n\in\mathbb{Z}^+$ then eq.(\ref{II.6}) becomes independent of $N$ as $N_\varphi$ becomes $\frac{1}{4}$. Hence, the inversion condition becomes $\frac{e^{2r}}{4}>n_P$ which is quite difficult as $r$ needs to be extremely high for population inversion to occur. Consider if $n_P\sim 10^9$, in such a scenario $r$ needs to be higher than or equal to 8.75. If $\varphi=(2n+1)\pi$, in such a scenario $N_\varphi$ becomes maximum. Now for a simple Bose-Einstein condensate, number of atoms achievable in a standard experimental set-up easily reaches as high as $10^7$ and even for very small squeezing, population inversion happens naturally. One can now easily solve eq.(\ref{II.6})
\begin{equation}\label{II.7}
n_G(t)=\mathcal{A}_G ~e^{\Gamma_{\text{GL}}(e^{2r}N_\varphi-n_P)t}-\frac{e^{2r}N_\varphi(n_P+1)}{e^{2r}N_\varphi-n_P}
\end{equation}
where the coefficient $\mathcal{A}_G$ is defined as $\mathcal{A}_G\equiv n_G(0)+\frac{e^{2r}N_\varphi(n_P+1)}{e^{2r}N_\varphi-n_P}$ with $n_G(0)$ denoting the number of gravitons in the system initially. If just one spontaneous emission happens $n_G(0)=1$, then $n_G(t)$ starts to grow exponentially leading to a coherent beam of gravitons which is equivalent to a LASER for photons.

\noindent If we now consider the case from \cite{ORM}, where $\omega=\omega_0+\omega_P$, we obtain the time evolution equation for the number of gravitons as 
\begin{equation}\label{II.NGL}
\frac{dn_G(t)}{dt}=-\Gamma_{\text{GL}} n_G(t)(n_P+N+1)+\Gamma_{\text{GL}} N
\end{equation}
indicating that graviton lasing is impossible in this process as there are no possible way for population inversion. As a result we need, for this moment, to rely on the Bose-Einstein condensate based population inversion mechanism described earlier where the resonance condition is $\omega_0=\omega+\omega_P$ (Case considered in this work). 

\noindent \textit{Experimental feasibility:}
In a real experimental scenario a simple system using a Bose-Einstein condensate can be designed. At first we need to investigate the coefficient $\Gamma_{\text{GL}}=\frac{\pi^2\hbar\omega q^2 G}{\varepsilon_0 c^2 m_0\omega_0\omega_P V^2}\delta(\omega+\omega_P-\omega_0)\rightarrow \frac{\pi\hbar\omega q^2 G\tau}{2\varepsilon_0 c^2 m_0\omega_0\omega_P V^2}$ where for a total measurement time the delta function is replaced by $\delta(\omega+\omega_P-\omega_o)\rightarrow \frac{\tau}{2\pi}$ and $\varepsilon_0$ gives the permittivity of free space. For $m_0\sim 10^{-25}$ kg, $V\sim 10^{-6}$ $\text{m}^3$, $\tau\sim 10^{5}$ sec, $\omega\simeq 1000$ Hz, $\omega_P\simeq 500$ Hz, $\omega_0\simeq 1500$ Hz, and $q\sim 1$ C, we obtain the numerical value of the coefficient $\Gamma_{\text{GL}}$ to be $\Gamma_{\text{GL}}\sim 10^{-10}$ Hz.
\begin{figure}[ht!]
\begin{center}
\includegraphics[scale=0.34]{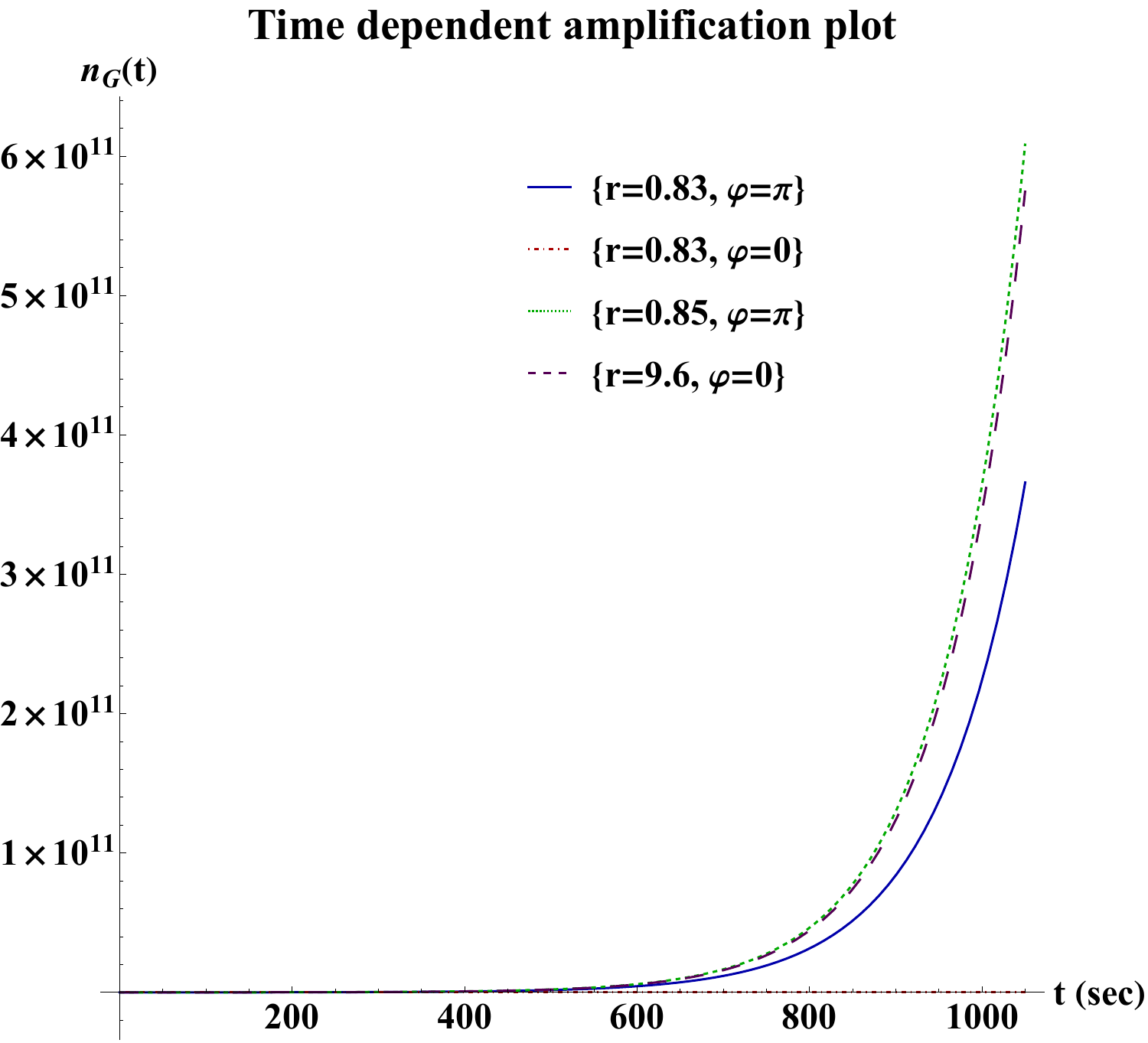}
\caption{The time dependent expectation of the graviton number operator is plotted against the total time. We observe that higher squeezing leads to significant amount gain in the graviton numbers over time and it is also dependent on the tuning angle $\varphi$.\label{Figure_OTM}}
\end{center}
\end{figure}
In Fig.(\ref{Figure_OTM}), we plot $n_G(t)$ against time for different values of the phonon squeezing parameter and effective tuning angle $\varphi$. If the phase $\theta$ corresponding to the coherent states are fixed then $\varphi$ can be tuned using the squeezing angle $\phi$. We find out that for effective angle $\varphi=0$, the population inversion is never achieved, however with $\varphi=\pi$ and a squeezing of only $r=0.83$, a significant gain in the number of gravitons in the quantum system is achieved indicating effective population inversion. Even we observe that for $r=0.85$ the gain is amplified corresponding to the $r=0.83$ case. Even for $\varphi=0$, we observe from the above figure that if the phonon is highly squeezed (here, $r=9.6$), the gain is increased significantly. This actually indicates towards a possible experimental implementation of the graviton laser model. Now, to gain true coherence among the emitted gravitons, we need to actually consider $\mathcal{N}$ as a time dependent variable instead of treating it as a bath which will let us move towards a more robust experimental set-up for such graviton laser model.

\begin{figure}
\begin{center}
\includegraphics[scale=0.9]{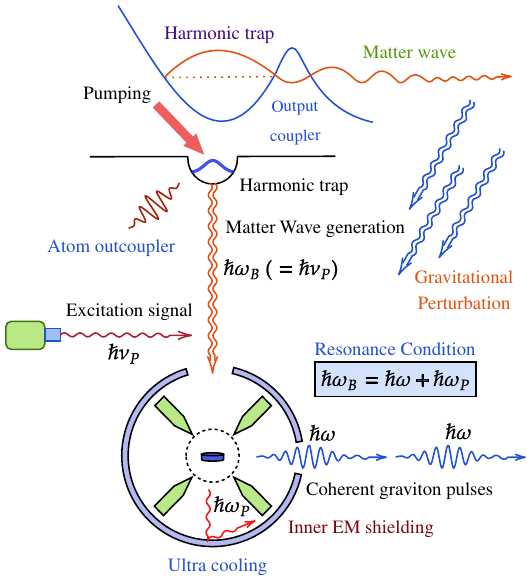}
\caption{A schematic diagram for graviton laser generation using ultra-cold bosonic systems in a Cavity-QED set up (Not to scale).\label{GRASER_OTM}}
\end{center}
\end{figure}

\textit{Experimental Proposal-} We propose a simple graviton laser generation device using ultra-cold bosonic systems. Systems like a liquid helium (instead of a Bose-Einstein condensate) can also work for our experimental set-up. The experimental proposal is depicted in Fig.(\ref{GRASER_OTM}). For the first step of our experimental model inside of a harmonic trap an ultra cold bosonic system is placed. The atoms are then outcoupled using atom outcouplers which results in the generation of a coherent matter wave beam with frequency $\hbar\omega_B$. An external electromagnetic excitation signal with each photons carrying an energy of $\hbar\nu_P$ is now focussed directly on the matter wave beam such that $\nu_P=\omega_B$. This results in an excitation of the bosons of the matter waves. A second optical tweezer or traps are then used to collect the matter wave beam with excited bosons. Using a second phase of cooling (Laser cooling for Bose-Einstein condensates) the excited matter wave is again super-cooled which results in the immediate deexcitation which produces photons and gravitons each carrying an energy $\hbar\omega_P$ and $\hbar\omega$ such that $\hbar\omega_B=\hbar\omega+\hbar\omega_P$. The inner surface of the optical tweezer set-up is covered with reflecting coating and the exit channel is covered completely with a polarizer to trap the emitted electromagnetic signal. If the optical-tweezer is designed with a atomic squeezer such that the phonon modes are squeezed then it results in population inversion and generation of coherent graviton pulses. Overtime the coherent graviton pulses become strong enough to be detected as a small gravitational perturbation which can be turned off immediately by switching off the excitation pulse resulting in no further deexcitation. As a result no more coherent graviton pulses are generated. This process needs to refined which we plan to do in a future endeavour.

\textit{Naturally occurring phenomena:} In our universe a natural graviton laser generation scenario may be observed for a Neutron star binary. For a neutron stars the density of the core is extremely high which may replicate highly squeezed quantum states with a extremely high number of particles. Now a binary Neutron star inspiral mimics an oscillating system. The electromagnetic signal from the counter-part star acts as an excitation signal resulting in a general set-up for graviton laser generation. In such a scenario, the binary inspiral should result in a natural generation device for graviton laser. It may be possible that the strong gravitational waves generated from the spiral actually is coherent graviton signals behaving like a gravitational wave.
\begin{figure}[ht!]
\begin{center}
\includegraphics[scale=1.0]{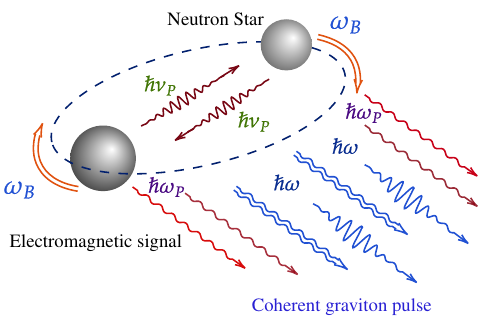}
\caption{A schematic diagram of graviton laser generation in a Neutron star binary system (Not to scale).\label{Neutron_Star_OTM}}
\end{center}
\end{figure}
In Fig.(\ref{Neutron_Star_OTM}), we show a schematic diagram of a Neutron star binary where the photons with energy $\hbar\nu_P$ works as the excitation signal. The emitted electromagnetic signal with energy $\hbar\omega_P$ represents the strong electromagnetic signal that we observe using high sensitive telescopes. Such a system then results in a generation of coherent gravitons with each carrying an energy of $\hbar\omega$.

\textit{Conclusion-} In this work, we extend our proposal for a true graviton detection model in \cite{ORM} by calculating the Lindblad master equation using the trilinear interaction Hamiltonian in eq.(\ref{II.1}). We start with the total density matrix of the system and by tracing out the photon as well as detector degrees of freedom which lands us on the Lindblad master equation involving the reduced density matrix corresponding to the graviton degrees of freedom. Now using the master equation and taking the expectation of the graviton number operator, we obtain the rate equation for graviton numbers in eq.(\ref{II.5}). We observe that true population inversion is possible if and only if we work with squeezed coherent states which is naturally achievable using ultra-cold atoms or a Bose-Einstein condensate. For a condensate based model, we then obtain the time dependence of the number of gravitons in the system and conclude that natural population inversion is possible for such a model. Unlike standard LASER theory, we don not require a three level system, rather the inversion occurs from the difference of the number of phonons (multiplied by the exponential squeezing factor) and the number of photons in the system. We have then plotted the graviton number against time and investigate whether the model is experimentally feasible. We find out that this novel model not only will serve as a step towards a truer graviton detection scenario but also may lead to actual generation of a graviton laser and it may fundamentally change our concept regarding graviton detection and the nature of gravitons. We then propose an actual experimental proposal for graviton laser generation in Fig.(\ref{GRASER_OTM}). We argue that  a two level ultra-cold bosonic system in a cavity-QED set up may allow for population inversion and generation of coherent graviton pulses. This simple set-up may lead to a more concrete experimental proposal for graviton laser generation in state of the art experimental laboratories in future. We also argue that for a cosmological scenario a Neutron star binary may be a natural source for graviton laser generation. We also expect to undergo a full analytical modelling of a binary neutron star spiral for graviton laser generation.

\onecolumngrid
\section{End Matter}
\subsection{Deriving the full master equation}
\noindent For the density matrix in the interaction picture $\hat{\rho}^I(t)$, the master equation after the Markov approximation reads
\begin{equation}\label{A.1}
\begin{split}
\frac{d\hat{\rho}^I(t)}{dt}=-\frac{i}{\hbar}\left[\hat{\mathcal{H}}^{\text{GL}}(t),\hat{\rho}^I(0)\right]-\frac{1}{\hbar^2}\int_0^t dt'[\hat{\mathcal{H}}^{\text{GL}}(t),[\hat{\mathcal{H}}^{\text{GL}}(t'),\hat{\rho}^I(t)]]
\end{split}
\end{equation}
The density matrix can be separated via implementing the Born approximation as $\hat{\rho}^I(t)=\hat{\rho}^{I}_{\text{GR}}(t)\otimes\hat{\rho}_P\otimes\hat{\rho}_{\text{UC}}$. Here UC in the subscript of $\hat{\rho}_{\text{UC}}$ denotes a detector with ultra-cold atoms. In order to obtain the master equation corresponding to the evolution of the Graviton part of the system, we need to take trace over the photon as well as detector degrees of freedom. This helps us to write down the master equation in a more simplified form
\begin{equation}\label{A.2}
\frac{d\hat{\rho}^I_{\text{GR}}(t)}{dt}=\text{tr}_{\text{UC}}\left[\frac{d\hat{\rho}^I(t)}{dt}\right]=-\frac{1}{\hbar^2}\int_0^tdt' \text{tr}_{\text{UC}}\left[\text{tr}_{\text{P}}\left[[\hat{\mathcal{H}}^{\text{GL}}(t),[\hat{\mathcal{H}}^{\text{GL}}(t'),\hat{\rho}^I(t)]]\right]\right]
\end{equation}
where the interaction Hamiltonian from eq.(\ref{II.3}) reads  $\hat{\mathcal{H}}^{\text{GL}}(t)=-i\hbar \mathcal{A}\left(\hat{b}^\dagger\hat{a}^\dagger\hat{\chi}\exp[i\varpi t]-\hat{b}\hat{a}\hat{\chi}^\dagger \exp[-i\varpi t]\right)$.
We now obtain the trace reduced $\hat{\kappa}_{\text{GR}}(t)=\text{tr}_{\text{UC}}\left[\text{tr}_{\text{P}}\left[[\hat{\mathcal{H}}^{\text{GL}}(t),[\hat{\mathcal{H}}^{\text{GL}}(t'),\hat{\rho}^I(t)]]\right]\right]$ commutator as
\begin{equation}\label{A.3}
\begin{split}
\hat{\kappa}_{\text{GR}}(t)=-\hbar^2\mathcal{A}^2\biggr[e^{i\varpi (t-t')}\Bigr[&-\hat{b}^\dagger\hat{b}\hat{\rho}^I_\text{GR}(t)\text{tr}_{\text{P}}\left[\hat{\rho}_P\hat{a}^\dagger\hat{a}\right]\text{tr}_{\text{UC}}\left[\hat{\rho}_\text{UC}\hat{\chi}\hat{\chi}^\dagger\right]+\hat{b}^\dagger\hat{\rho}^I_\text{GR}(t)\hat{b}\text{tr}_{\text{P}}\left[\hat{\rho}_P\hat{a}\hat{a}^\dagger\right]\text{tr}_{\text{UC}}\left[\hat{\rho}_\text{UC}\hat{\chi}^\dagger\hat{\chi}\right]\\&+\hat{b}\hat{\rho}^I_\text{GR}(t)\hat{b}^\dagger\text{tr}_{\text{P}}\left[\hat{\rho}_P\hat{a}^\dagger\hat{a}\right]\text{tr}_{\text{UC}}\left[\hat{\rho}_\text{UC}\hat{\chi}\hat{\chi}^\dagger\right]-\hat{\rho}^I_\text{GR}(t)\hat{b}\hat{b}^\dagger\text{tr}_{\text{P}}\left[\hat{\rho}_P\hat{a}\hat{a}^\dagger\right]\text{tr}_{\text{UC}}\left[\hat{\rho}_\text{UC}\hat{\chi}^\dagger\hat{\chi}\right]\Bigr]\\
+e^{-i\varpi (t-t')}\Bigr[&-\hat{b}\hat{b}^\dagger\hat{\rho}^I_\text{GR}(t)\text{tr}_{\text{P}}\left[\hat{\rho}_P\hat{a}\hat{a}^\dagger\right]\text{tr}_{\text{UC}}\left[\hat{\rho}_\text{UC}\hat{\chi}^\dagger\hat{\chi}\right]+\hat{b}\hat{\rho}^I_\text{GR}(t)\hat{b}^\dagger\text{tr}_{\text{P}}\left[\hat{\rho}_P\hat{a}^\dagger\hat{a}\right]\text{tr}_{\text{UC}}\left[\hat{\rho}_\text{UC}\hat{\chi}\hat{\chi}^\dagger\right]\\&+\hat{b}^\dagger\hat{\rho}^I_\text{GR}(t)\hat{b}\text{tr}_{\text{P}}\left[\hat{\rho}_P\hat{a}\hat{a}^\dagger\right]\text{tr}_{\text{UC}}\left[\hat{\rho}_\text{UC}\hat{\chi}^\dagger\hat{\chi}\right]-\hat{\rho}^I_\text{GR}(t)\hat{b}^\dagger\hat{b}\text{tr}_{\text{P}}\left[\hat{\rho}_P\hat{a}^\dagger\hat{a}\right]\text{tr}_{\text{UC}}\left[\hat{\rho}_\text{UC}\hat{\chi}\hat{\chi}^\dagger\right]\Bigr]
\biggr]
\end{split}
\end{equation}
where we have made use of the cyclic property of trace. Now $\hat{\rho}_P=|n_P\rangle\langle n_P|$, and as a result, we obtain the trace values of $\text{tr}_{\text{P}}\left[\hat{\rho}_P\hat{a}\hat{a}^\dagger\right]$ and $\text{tr}_{\text{P}}\left[\hat{\rho}_P\hat{a}^\dagger\hat{a}\right]$ as
\begin{equation}\label{A.4}
\text{tr}_{\text{P}}\left[\hat{\rho}_P\hat{a}\hat{a}^\dagger\right]=n_P+1~\text{and}~~\text{tr}_{\text{P}}\left[\hat{\rho}_P\hat{a}^\dagger\hat{a}\right]=n_P~.
\end{equation}
Similarly, we obtain for the ultra-cold sector as
\begin{equation}\label{A.5}
\text{tr}_{\text{UC}}\left[\hat{\rho}_\text{UC}\hat{\chi}\hat{\chi}^\dagger\right]=\mathcal{N}+1~\text{and}~~\text{tr}_{\text{UC}}\left[\hat{\rho}_\text{UC}\hat{\chi}^\dagger\hat{\chi}\right]=\mathcal{N}~.
\end{equation}
We now need to calculate the integrals $\int_0^t dt' e^{\pm i\varpi(t-t')}$. We define a new time parameter $\tau\equiv t-t'$ and the integrals can be recast as $\int_0^t dt' e^{\pm i\varpi(t-t')}\rightarrow \int_0^t d\tau e^{\pm i\varpi \tau}$. We now implement the well-known infinite time approximation which lets us write down the integrals as $\mathcal{I}_\pm=\int_0^\infty d\tau e^{\pm i\varpi \tau}$. We can further replace the integrals by the symmetric values as $\mathcal{I}^\infty_\pm=\frac{1}{2}\int_{-\infty}^\infty d\tau e^{\pm i\varpi \tau}$. The integral can be obtained as
\begin{equation}\label{A.6}
\mathcal{I}_\pm=\frac{1}{2}\int_{-\infty}^\infty d\tau e^{\pm i\varpi \tau}=\frac{1}{2}2\pi\delta(\varpi)=\pi\delta(\varpi)~.
\end{equation}
We can then obtain simply the transition rate factor as $\Gamma=\frac{1}{\hbar^2}(\hbar^2\mathcal{A})^2\mathcal{I}_\pm=\pi \mathcal{A}^2\delta(\omega+\omega_P-\omega_0)=\frac{\Gamma_{\text{GL}}}{2}$. We can then substitute eq.(s)(\ref{A.3},\ref{A.4},\ref{A.5},\ref{A.6}) in eq.(\ref{A.2}), and obtain the final form of the master equation as
\begin{equation}\label{A.7}
\frac{d\hat{\rho}^I_{\text{GR}}(t)}{dt}=-\Gamma_{\text{GL}}\left[(n_P+1)\mathcal{N}\left[\hat{b}^\dagger \hat{\rho}^I_{\text{GR}}(t)\hat{b}-\frac{1}{2}\{\hat{b}\hat{b}^\dagger,\hat{\rho}^I_{\text{GR}}(t)\}\right]\right]-\Gamma_{\text{GL}}\left[n_P(\mathcal{N}+1)\left[\hat{b} \hat{\rho}^I_{\text{GR}}(t)\hat{b}^\dagger-\frac{1}{2}\{\hat{b}^\dagger\hat{b},\hat{\rho}^I_{\text{GR}}(t)\}\right]\right]
\end{equation}
which is eq.(\ref{II.4}) in our manuscript.
\end{document}